\documentclass[a4paper,11pt]{article}
\usepackage{jheppub}
\usepackage[T1]{fontenc}
\usepackage[utf8]{inputenc}
\usepackage{lmodern}
\usepackage{microtype}
\usepackage{amsmath,amssymb,amsthm,mathtools,bm}
\usepackage{booktabs}
\usepackage{enumitem}
\usepackage{graphicx}

\newcommand{\dd}{\mathrm{d}}
\newcommand{\ii}{\mathrm{i}}
\newcommand{\ee}{\mathrm{e}}

\newcommand{\Real}{\operatorname{Re}}
\newcommand{\Imag}{\operatorname{Im}}

\newcommand{\om}{\omega}
\newcommand{\Om}{\Omega}

\newcommand{\cO}{\mathcal{O}}
\newcommand{\intk}{\int\!\frac{\dd k}{2\pi}}
\newcommand{\abs}[1]{\left|#1\right|}
\newcommand{\avg}[1]{\left\langle#1\right\rangle}
\graphicspath{{figures/}}

\title{\boldmath Transient Coherence and Multiscale Spectral Structure after a Mass Quench in a 1+1-Dimensional Scalar Field}

\author[a,b]{Gabriel Abell\'an\note{Corresponding author.}}
\author[a,b]{and Nelson Bol\'ivar}

\affiliation[a]{Departamento de F\'isica, Facultad de Ciencias, Universidad Central de Venezuela,\
Av. Los Ilustres, Caracas 1041-A, Venezuela}
\affiliation[b]{Astrum Drive Technologies,\
Dallas Pkwy Unit 120 B, Frisco, TX 75034, USA}

\emailAdd{gabriel.abellan@ciens.ucv.ve}
\emailAdd{nelson.bolivar@ciens.ucv.ve}

\abstract{A mass quench generates anomalous pair correlations that are absent from the final particle occupations. For a free real scalar field in $1+1$ dimensions, we define a bilocal observable by smearing the statistical Keldysh correlator with a Morlet kernel on the finite post-quench relative-time interval. Its excess over the dephased correlator gives a dimensionless relative coherence weight. At fixed momentum, this weight is exactly independent of the initial temperature for Gaussian thermal states evolved with the same free Hamiltonian: the common thermal amplitude cancels between numerator and denominator. The cancellation is not a purity criterion and is not exclusive to wavelets. We derive the finite-window response and show that the anomalous signal depends essentially on the observation window; an admissible zero-mean wavelet annihilates the relative-time-constant anomalous term on the full line. We compare this response with the signed finite-time Wigner spectrum and derive a distinct Gaussian-extension positivity bound. Exact smooth-quench occupations provide an independent dynamical reference. Finally, we formulate a renormalized Hartree extension and delimit the oscillatory and nonoscillatory regimes of a scalar mode on a fixed two-dimensional de Sitter background.}


\begin{document}
\maketitle
\flushbottom

\section{Introduction}

This work defines a finite-window, scale-resolved diagnostic of anomalous pair coherence in the statistical Keldysh correlator of a quenched scalar field. Its normalized response is exactly independent of the initial temperature at fixed momentum in the free Gaussian problem. Its sensitivity to coherence, however, depends on the relative-time window: a zero-mean wavelet removes the constant anomalous term on an unbounded relative-time domain. Establishing both properties is necessary to interpret a multiscale map as a physical diagnostic rather than as an image of the correlator.

Nonequilibrium quantum field theory is naturally expressed through two-time correlation functions on the closed time contour~\cite{Kamenev,CalzettaHu,Berges}. Time-translation invariance in equilibrium reduces these functions to a dependence on $t_1-t_2$. A quench breaks this symmetry: average time $T=(t_1+t_2)/2$ tracks state evolution, whereas relative time $\tau=t_1-t_2$ resolves its oscillatory structure. Occupations in the final particle basis do not determine the anomalous expectation values or their phase. This does not mean that every equal-time measurement loses coherence: the complete field--momentum covariance matrix also contains it.

The finite-time Wigner transform retains this information in a signed frequency representation. Negative lobes do not invalidate the representation or imply a negative probability. A Gaussian filter of an extended post-quench kernel can obey a mode-dependent positivity bound, but this is a different object from the exact finite-window transform. A complex wavelet transform is likewise not a positive probability density and does not remove this distinction. Its purpose here is to specify a bilocal smearing with a chosen relative-time position and scale.

Exact correlators and generalized stationary descriptions of scalar mass quenches have been studied in two and higher dimensions~\cite{Mandal2015,Banerjee2019}; quantum quenches in integrable field theory provide a related framework~\cite{Fioretto:2009yq}. These are distinct from the mechanism of thermalization in generic isolated systems~\cite{Rigol:2007juv}. Smooth and instantaneous quenches were compared in Refs.~\cite{Das:2014hqa,Das2015}. Keldysh methods also describe quenched oscillator models and local scalar perturbations~\cite{Choudhury:2022ati,Ageev:2022kpm,RadovskayaSemenov}. The occupation and mode solutions used below are established results, not claims of novelty.

Wavelets have been used for field-theory regularization, scale-dependent Green functions, and Hamiltonian multiresolution representations~\cite{Halliday:1994fe,Federbush:1995rt,Bulut:2013bg,AltaiskyKaputkina,Brennen:2014iqu,GeorgeWavelets,Basak:2026ioz,Basak:2026meu}. Our construction instead smears the relative-time dependence of a bilocal statistical observable. The contribution is the explicit finite-window response, its dephased subtraction, and a quantitative comparison with the Wigner response under stated observation conditions. The thermal cancellation follows from linearity and a common mode amplitude; it is a useful consistency property, not a wavelet-specific protection principle.

We first derive the sudden-quench correlators, then construct the finite-window Wigner and Morlet responses and their normalized excess weights. Smooth quenches supply an exact interpolation to adiabatic evolution. The final sections separate infrared and ultraviolet issues from two extensions: a self-consistent Gaussian interaction approximation and a prescribed expanding geometry. We use $\hbar=c=1$, metric signature $(-,+)$, and positive $m_i,m_f$ unless an infrared prescription is explicitly stated.

\section{Scalar field with a time-dependent mass}

Consider a real scalar field in one spatial dimension with action
\begin{equation}
 S[\phi]=\frac{1}{2}\int\dd t\,\dd x\,
 \left[(\partial_t\phi)^2-(\partial_x\phi)^2-m^2(t)\phi^2\right].
 \label{eq:action}
\end{equation}
The equation of motion is
\begin{equation}
  \left[\partial_t^2-\partial_x^2+m^2(t)\right]\phi(t,x)=0.
 \label{eq:kg-minkowski}
\end{equation}
For a spatially homogeneous mass, momentum remains a good quantum number. We expand
\begin{equation}
  \phi(t,x)=\intk\left[a_k u_k(t)\ee^{\ii kx}
  +a_k^{\dagger}u_k^*(t)\ee^{-\ii kx}\right],
 \label{eq:mode-expansion}
\end{equation}
with
\begin{equation}
  [a_k,a_q^{\dagger}]=2\pi\delta(k-q).
\end{equation}
Each mode obeys
\begin{equation}
 \ddot u_k(t)+\om_k^2(t)u_k(t)=0,
 \qquad \om_k^2(t)=k^2+m^2(t).
 \label{eq:mode-equation}
\end{equation}
Canonical quantization imposes the Wronskian condition
\begin{equation}
 u_k\dot u_k^*-\dot u_k u_k^*=\ii.
 \label{eq:wronskian}
\end{equation}
This condition is the mode-by-mode form of the canonical commutation relations. It is also the most useful invariant for following the dynamics through a time-dependent background.

\section{The sudden quench}
\label{sec:sudden}

We begin with
\begin{equation}
 m^2(t)=m_i^2\Theta(-t)+m_f^2\Theta(t),
 \label{eq:sudden-mass}
\end{equation}
where the initial state is the vacuum associated with $m_i$. Define
\begin{equation}
 \om_{i,k}=\sqrt{k^2+m_i^2},
 \qquad
 \om_{f,k}=\sqrt{k^2+m_f^2}.
 \label{eq:frequencies}
\end{equation}
For $t<0$, the positive-frequency mode is
\begin{equation}
 u_{i,k}(t)=\frac{\ee^{-\ii\om_{i,k}t}}{\sqrt{2\om_{i,k}}}.
 \label{eq:initial-mode}
\end{equation}
For $t>0$, write the same solution in the final basis,
\begin{equation}
 u_k(t)=\frac{1}{\sqrt{2\om_{f,k}}}
 \left(\alpha_k\ee^{-\ii\om_{f,k}t}+\beta_k\ee^{+\ii\om_{f,k}t}\right).
 \label{eq:post-mode}
\end{equation}
Since the mass changes by a finite step and contains no delta function, both $u_k$ and $\dot u_k$ are continuous at $t=0$. Matching gives
\begin{equation}
 \alpha_k+\beta_k=\sqrt{\frac{\om_{f,k}}{\om_{i,k}}},
 \qquad
 \alpha_k-\beta_k=\sqrt{\frac{\om_{i,k}}{\om_{f,k}}},
 \label{eq:matching}
\end{equation}
and therefore
\begin{equation}
 \alpha_k=\frac{\om_{f,k}+\om_{i,k}}
 {2\sqrt{\om_{f,k}\om_{i,k}}},
 \qquad
 \beta_k=\frac{\om_{f,k}-\om_{i,k}}
 {2\sqrt{\om_{f,k}\om_{i,k}}}.
 \label{eq:bogoliubov}
\end{equation}
The coefficients are real for the instantaneous quench with the phase convention in Eq.~\eqref{eq:initial-mode}. Their canonical identity is
\begin{equation}
 \abs{\alpha_k}^2-\abs{\beta_k}^2=1.
 \label{eq:unitarity}
\end{equation}
The number of final particles in the initial vacuum is
\begin{equation}
 n_k=\abs{\beta_k}^2
 =\frac{(\om_{f,k}-\om_{i,k})^2}{4\om_{f,k}\om_{i,k}}.
 \label{eq:occupation}
\end{equation}
This occupation is not thermal. It is the consequence of expressing the initial vacuum in the final quasiparticle basis. The same transformation also produces anomalous correlations, which are responsible for the explicit average-time dependence below.

\subsection{High-momentum behavior}

For $k\to\infty$,
\begin{equation}
 \beta_k=\frac{m_f^2-m_i^2}{4k^2}+\cO(k^{-4}),
 \qquad
 n_k=\frac{(m_f^2-m_i^2)^2}{16k^4}+\cO(k^{-6}).
 \label{eq:high-k-beta}
\end{equation}
The occupation itself is ultraviolet soft. In $1+1$ dimensions this softness is sufficient to make the particle-production contribution to the excitation energy convergent. The dimension dependence of this statement is discussed in Sec.~\ref{sec:uv}.

\section{Keldysh correlators}
\label{sec:keldysh}

For the initial vacuum, the mode Wightman functions are
\begin{equation}
 G_k^>(t_1,t_2)=u_k(t_1)u_k^*(t_2),
 \qquad
 G_k^<(t_1,t_2)=u_k(t_2)u_k^*(t_1).
 \label{eq:wightman}
\end{equation}
We define the statistical and spectral correlators by
\begin{equation}
 F_k(t_1,t_2)=\frac{1}{2}\left[G_k^>(t_1,t_2)+G_k^<(t_1,t_2)\right],
 \label{eq:statistical}
\end{equation}
\begin{equation}
 \rho_k(t_1,t_2)=\ii\left[G_k^>(t_1,t_2)-G_k^<(t_1,t_2)\right].
 \label{eq:spectral}
\end{equation}
The statistical correlator depends on the state, whereas the spectral correlator is fixed by the canonical commutator. With the convention $G_k^K=-2\ii F_k$, the Keldysh propagator is directly proportional to the statistical function.

Introduce average and relative times,
\begin{equation}
 T=\frac{t_1+t_2}{2},
 \qquad
 \tau=t_1-t_2.
 \label{eq:average-relative}
\end{equation}
In the post-quench sector $t_1,t_2>0$, substitute Eq.~\eqref{eq:post-mode} into Eqs.~\eqref{eq:statistical}--\eqref{eq:spectral}. One obtains
\begin{equation}
 F_{k,++}(T,\tau)=A_k\cos(\om_{f,k}\tau)+B_k\cos(2\om_{f,k}T),
 \label{eq:F-post}
\end{equation}
where
\begin{equation}
 A_k=\frac{\om_{f,k}^2+\om_{i,k}^2}
 {4\om_{i,k}\om_{f,k}^2},
 \qquad
 B_k=\frac{\om_{f,k}^2-\om_{i,k}^2}
 {4\om_{i,k}\om_{f,k}^2}.
 \label{eq:AB}
\end{equation}
The spectral function is
\begin{equation}
 \rho_{k,++}(T,\tau)=\frac{\sin(\om_{f,k}\tau)}{\om_{f,k}}.
 \label{eq:rho-post}
\end{equation}
Equation~\eqref{eq:F-post} exhibits the central separation of physical information. The first term is stationary in average time and oscillates at the final frequency in relative time. The second term is independent of relative time but oscillates at twice the final frequency in average time. It is the anomalous, or squeezed, coherence generated by the quench. It disappears when $m_i=m_f$.

The absence of an analogous average-time term in Eq.~\eqref{eq:rho-post} is not accidental. The spectral function measures the commutator and therefore does not encode the occupation of the state. The quench changes the statistical sector while preserving the canonical spectral structure.

\section{Finite-time Wigner representation}
\label{sec:wigner}

For a stationary correlator, the Fourier transform in relative time is the frequency spectrum. In a nonstationary problem, the natural generalization is the Wigner transform,
\begin{equation}
 F_{W,k}(T,\Om)=\int\dd\tau\,\ee^{\ii\Om\tau}F_k(T,\tau).
 \label{eq:wigner-general}
\end{equation}
For the post-quench sector, the conditions $t_1,t_2>0$ restrict the relative time to $-2T<\tau<2T$ when $T>0$. Thus
\begin{equation}
 F_{W,k}^{++}(T,\Om)=\int_{-2T}^{2T}\dd\tau\,\ee^{\ii\Om\tau}F_{k,++}(T,\tau).
 \label{eq:wigner-finite}
\end{equation}
Using Eq.~\eqref{eq:F-post},
\begin{align}
 F_{W,k}^{++}(T,\Om)
 ={}&A_k\left[
 \frac{\sin[2T(\Om+\om_{f,k})]}{\Om+\om_{f,k}}
 +\frac{\sin[2T(\Om-\om_{f,k})]}{\Om-\om_{f,k}}
 \right]
 \nonumber\\
 &+2B_k\cos(2\om_{f,k}T)\frac{\sin(2T\Om)}{\Om}.
 \label{eq:wigner-post}
\end{align}
The apparent singularities are removable, with $\lim_{x\to0}\sin(2Tx)/x=2T$. The first line produces peaks near $\Om=\pm\om_{f,k}$, with width of order $1/T$. The second line is centered at zero relative-time frequency and carries the average-time coherence. Thus the Wigner transform distinguishes the final quasiparticle spectrum from the memory of the quench.

The finite integration interval is physically meaningful. A finite average time does not provide infinite relative-time resolution, so the Wigner peaks are broadened. This broadening is not a failure of the transform; it is the time-frequency uncertainty inherent in the nonstationary problem.

\subsection{Gaussian-filtered relative-time spectrum}
\label{sec:gaussian-filter}

A distinct analysis object is obtained by extending the algebraic expression in Eq.~\eqref{eq:F-post} to all $\tau$ and applying a Gaussian envelope:
\begin{align}
 F_{G,k}(T,\Om)
 &=\int_{-\infty}^{\infty}\dd\tau\,\ee^{\ii\Om\tau-\tau^2/(2a^2)}
 \left[A_k\cos(\om_{f,k}\tau)+B_k\cos(2\om_{f,k}T)\right]\nonumber\\
 &=\sqrt{2\pi}a\,\ee^{-a^2\Om^2/2}
 \left[A_k\ee^{-a^2\om_{f,k}^2/2}\cosh(a^2\Om\om_{f,k})
 +B_k\cos(2\om_{f,k}T)\right].
 \label{eq:gaussian-wigner}
\end{align}
This extension is not the actual crossed-sector correlator, nor is it Gaussian filtering of the hard-window spectrum in Eq.~\eqref{eq:wigner-finite}. The distinction matters near the quench, where $a$ need not be small compared with $2T$.

For $B_k\ne0$ and $a>0$, Eq.~\eqref{eq:gaussian-wigner} is nonnegative for all $T$ and $\Om$ if and only if
\begin{equation}
 a\leq a_{\rm pos,k}
 =\frac{\sqrt{2\ln(A_k/\abs{B_k})}}{\om_{f,k}}.
 \label{eq:gaussian-positivity}
\end{equation}
Sufficiency follows from $\cosh x\geq1$. For necessity, set $\Om=0$
and choose $\cos(2\om_{f,k}T)=-\operatorname{sgn}B_k$.
Equality gives a zero at that point; strict inequality gives strict
positivity at every finite frequency. If $B_k=0$, there is no finite
upper bound on $a$.

Writing $D=\abs{m_f^2-m_i^2}$, the ultraviolet asymptote is
\begin{equation}
 a_{\rm pos,k}\sim
 \frac{\sqrt{2\ln(2k^2/D)}}{\abs{k}}\longrightarrow0.
 \label{eq:apos-uv}
\end{equation}
The logarithm has a dimensionless argument. No fixed $a>0$ ensures positivity uniformly in all momentum modes, but this fact alone does not prove that a momentum-integrated spectrum is negative. At small $k$, with $S=m_i^2+m_f^2$,
\begin{equation}
 a_{\rm pos,k}^2=
 \frac{2\ln(S/D)}{m_f^2}
 +\frac{2}{m_f^4}\left[\frac{2m_f^2}{S}-\ln(S/D)\right]k^2
 +\cO(k^4).
 \label{eq:apos-ir}
\end{equation}
The slope vanishes at $k=0$ for positive masses, while the sign of the curvature depends on the mass ratio. It is positive for the illustrated $m_i=1$, $m_f=2$ quench, not for every quench.

\begin{figure}[t]
 \centering
 \includegraphics[width=\linewidth]{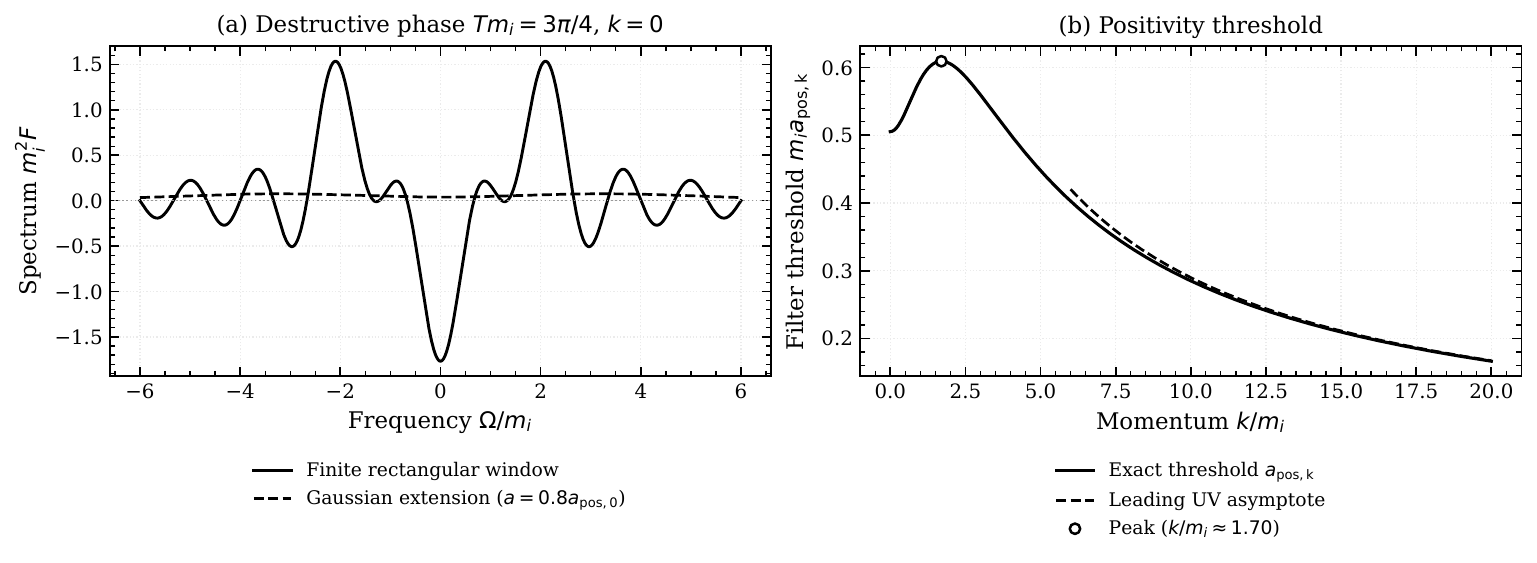}
 \caption{Spectral prescriptions for $m_f/m_i=2$. (a) Finite-window Wigner spectrum and Gaussian extension at $k=0$, $Tm_i=3\pi/4$, where $\cos(2m_fT)=-1$. The chosen width is $a=0.8a_{\rm pos,0}$, with $m_i a_{\rm pos,0}\simeq 0.505$ and $m_i a\simeq 0.404$, calculated from Eq.~\eqref{eq:gaussian-positivity}. (b) Exact threshold and leading ultraviolet asymptote, the latter shown only for $k/m_i\geq6$. The maximum is at $k/m_i\simeq1.70$, calculated from Eq.~\eqref{eq:gaussian-positivity}. The initial rise is specific to this mass ratio.}
 \label{fig:wigner-gaussian}
\end{figure}

Figure~\ref{fig:wigner-gaussian} compares the two observation prescriptions at the exactly destructive phase $T=3\pi/(2m_f)$ and $k=0$. It shows that smoothing changes the observable as well as its sign structure. The right panel displays the nonmonotonic threshold and its ultraviolet decrease; it is not a positivity theorem for the finite-window Wigner transform.

\subsection{Crossed sectors and Abel regularization}

The global transform at fixed $T>0$ samples three time-order sectors:
\begin{equation}
 \tau<-2T:\ -+,
 \qquad
 -2T<\tau<2T:\ ++,
 \qquad
 \tau>2T:\ + -.
 \label{eq:sectors}
\end{equation}
The crossed sectors contain persistent oscillations, so their full-line
Fourier transform is distributional. An Abel factor provides one convenient
representation of this transform,
\begin{equation}
 F_{W,k,\epsilon}(T,\Om)=\int_{-\infty}^{\infty}\dd\tau\,
 \ee^{\ii\Om\tau}\ee^{-\epsilon\abs{\tau}}F_k(T,\tau),
 \qquad \epsilon>0.
 \label{eq:abel}
\end{equation}
For $\tau>2T$, the crossed correlator can be written as
\begin{align}
 F_{k,+-}(T,\tau)=\frac{1}{2\sqrt{\om_{f,k}\om_{i,k}}}
 \bigg[&\alpha_k\cos\left((\om_{f,k}-\om_{i,k})T
 +\frac{\om_{f,k}+\om_{i,k}}{2}\tau\right)
 \nonumber\\
 &+\beta_k\cos\left((\om_{f,k}+\om_{i,k})T
 +\frac{\om_{f,k}-\om_{i,k}}{2}\tau\right)\bigg].
 \label{eq:crossed}
\end{align}
The Abel limit is regular at generic frequency but becomes distributional at the crossed-sector frequencies
\begin{equation}
 \Om=\pm\frac{\om_{f,k}+\om_{i,k}}{2},
 \qquad
 \Om=\pm\frac{\om_{f,k}-\om_{i,k}}{2}.
 \label{eq:crossed-resonances}
\end{equation}
The regulator defines the Fourier transform as a distribution in the limit $\epsilon\to0^+$. At finite $\epsilon$ it is an imposed observation envelope, not a physical damping rate.

\section{Wavelet representation}
\label{sec:wavelets}

A continuous wavelet representation adds localization in relative time and scale~\cite{TorrenceCompo,Daubechies}. At fixed $T>0$, define
\begin{equation}
 W_{F,k}^{++}(T;b,s)=\frac{1}{\sqrt{s}}\int_{-2T}^{2T}\dd\tau\,
 F_{k,++}(T,\tau)\psi^*\!\left(\frac{\tau-b}{s}\right),\qquad s>0.
 \label{eq:cwt}
\end{equation}
Here $b$ is a relative-time position, not the physical average time. The observation interval changes with $T$. Consequently, even a stationary underlying kernel can have a $T$-dependent finite-window transform.

We first use the uncorrected Morlet kernel
\begin{equation}
 \psi_{\rm M}(x)=\pi^{-1/4}\ee^{-x^2/2}\ee^{\ii\omega_0x},
 \qquad \omega_0=6,
 \label{eq:morlet}
\end{equation}
and distinguish it explicitly from its zero-mean version below. The nominal scale for a frequency $\om_*$ is
\begin{equation}
 s_0=\frac{\omega_0}{\om_*}.
 \label{eq:wavelet-scale}
\end{equation}
It is the center of the Gaussian frequency envelope, not the exact maximum of the coefficient magnitude. For a single $\ee^{\ii\om_*\tau}$ on the full line, the factor $\sqrt{s}$ shifts that maximum to
\begin{equation}
 s_{\rm peak}=\frac{\omega_0+\sqrt{\omega_0^2+2}}{2\om_*}.
 \label{eq:exact-scale}
\end{equation}
Finite endpoints, multiple frequencies, and the chosen scale weighting can shift it further.

\subsection{Exact finite-window kernel}

Introduce the elementary response
\begin{equation}
 J_T(\om;b,s)=\frac{\pi^{-1/4}}{\sqrt{s}}
 \int_{-2T}^{2T}\dd\tau\,
 \ee^{-(\tau-b)^2/(2s^2)}
 \ee^{-\ii\omega_0(\tau-b)/s}\ee^{\ii\om\tau}.
 \label{eq:J-definition}
\end{equation}
Completing the square gives an exact expression in complex error functions:
\begin{align}
 d&=s\om-\omega_0,\qquad
 z_\pm=\frac{\pm2T-b}{\sqrt{2}s}-\frac{\ii d}{\sqrt{2}},\nonumber\\
 J_T(\om;b,s)
 &=\pi^{-1/4}\sqrt{\frac{\pi s}{2}}\,
 \ee^{\ii\om b-d^2/2}
 \left[\operatorname{erf}z_+-\operatorname{erf}z_-\right].
 \label{eq:J-finite}
\end{align}
The statistical and spectral maps therefore follow without a mode-evolution approximation:
\begin{align}
 W_{F,k}^{++}
 &=\frac{A_k}{2}\left[J_T(\om_{f,k})+J_T(-\om_{f,k})\right]
   +B_k\cos(2\om_{f,k}T)J_T(0),\label{eq:finite-WF}\\
 W_{\rho,k}^{++}
 &=\frac{J_T(\om_{f,k})-J_T(-\om_{f,k})}{2\ii\om_{f,k}}.
 \label{eq:spectral-wavelet}
\end{align}
Arguments $(b,s)$ are suppressed only on the right-hand sides. The signs of the phases in Eq.~\eqref{eq:J-finite} follow from the conjugated analyzing kernel in Eq.~\eqref{eq:cwt}.

For comparison, extending the diagonal kernel to the full line gives
\begin{align}
 W_{A,k}^{\infty}(b,s)
 =\frac{A_k}{2}\sqrt{2\pi}\pi^{-1/4}\sqrt{s}
 \big[&\ee^{-(s\om_{f,k}-\omega_0)^2/2}\ee^{+\ii\om_{f,k}b}\nonumber\\
 &+\ee^{-(s\om_{f,k}+\omega_0)^2/2}\ee^{-\ii\om_{f,k}b}\big].
 \label{eq:morlet-closed-form}
\end{align}
This is a useful limiting check, not a substitute for the finite response.

\subsection{Wavelets in the Keldysh basis}
\label{sec:wavelet-keldysh}

Linearity gives the connection to Keldysh theory directly:
\begin{equation}
 G_k^K=-2\ii F_k,\qquad W_{G^K,k}^{++}=-2\ii W_{F,k}^{++}.
 \label{eq:keldysh-wavelet}
\end{equation}
The spectral coefficients in Eq.~\eqref{eq:spectral-wavelet} remain state independent for a fixed free Hamiltonian. They are not independent of $T$ when the interval $[-2T,2T]$ changes. This separates a kinematic window effect from state-dependent average-time oscillations.

To make the operator normalization explicit, use a periodic box of length $L$ and
$\hat\phi_k=L^{-1/2}\int_0^L\dd x\,\ee^{-\ii kx}\hat\phi(x)$, with $k=2\pi n/L$.
Then $\hat\phi_k^\dagger=\hat\phi_{-k}$ and
$\frac12\langle\{\hat\phi_k(t_1),\hat\phi_{-k}(t_2)\}\rangle=F_k(t_1,t_2)$.
In infinite volume, $F_k$ instead denotes the coefficient after factoring out the momentum delta function.

\subsection{A wavelet-smeared Keldysh observable}
\label{sec:wavelet-observable}

With $\psi_{b,s}(\tau)=s^{-1/2}\psi((\tau-b)/s)$, define
\begin{equation}
 \widehat{\mathcal O}_{k,\psi}(T;b,s)=
 \frac12\int_{-2T}^{2T}\dd\tau\,\psi_{b,s}^*(\tau)
 \left\{\hat\phi_k\!\left(T+\frac{\tau}{2}\right),
 \hat\phi_{-k}\!\left(T-\frac{\tau}{2}\right)\right\}.
 \label{eq:wavelet-operator}
\end{equation}
Its expectation value is
\begin{equation}
 \mathcal K_{k,\psi}(T;b,s)=
 \langle\widehat{\mathcal O}_{k,\psi}(T;b,s)\rangle
 =W_{F,k}^{++}(T;b,s),\qquad
 \mathcal G^K_{k,\psi}=-2\ii\mathcal K_{k,\psi}.
 \label{eq:keldysh-observable}
\end{equation}
For a complex kernel, the Hermitian observables
$(\widehat{\mathcal O}+\widehat{\mathcal O}^\dagger)/2$ and
$(\widehat{\mathcal O}-\widehat{\mathcal O}^\dagger)/(2\ii)$ have expectation
values $\Real\mathcal K$ and $\Imag\mathcal K$. This is a bilocal correlation
measurement, not a positive single-time number operator.

The dephased reference is
\begin{equation}
 F_k^{\rm diag}(\tau)=A_k\cos(\om_{f,k}\tau).
 \label{eq:dephased-correlator}
\end{equation}
It is the average-time-dephased two-point function. Its wavelet transform must use the same finite interval as the full correlator. Thus
\begin{equation}
 \Delta\mathcal K_{k,\psi}
 =W_{F,k}^{++}(T;b,s)-W_{F,k}^{\rm diag,++}(T;b,s)
 =B_k\cos(2\om_{f,k}T)J_T(0;b,s).
 \label{eq:wavelet-coherence-coefficient}
\end{equation}
For an observation region $\mathcal B$, define the relative coherence weight
\begin{equation}
 \mathfrak C_{k,\psi}(T;\mathcal B)=
 \frac{\displaystyle\int_{\mathcal B}\frac{\dd b\,\dd s}{s^2}
 \abs{\Delta\mathcal K_{k,\psi}}^2}
 {\displaystyle\int_{\mathcal B}\frac{\dd b\,\dd s}{s^2}
 \abs{\mathcal K_{k,\psi}}^2},
 \label{eq:wavelet-coherence-fraction}
\end{equation}
provided the denominator is nonzero. We use
$\mathcal B(T)=[-T,T]\times[s_0/2,2s_0]$, with $s_0=\omega_0/\om_{f,k}$, for the figures.
The ratio is nonnegative, but the diagonal and anomalous coefficients need not be orthogonal on $\mathcal B$. It is therefore not generally bounded by one and should not be interpreted as a probability fraction. Both the response $J_T(0)$ and the integration region depend on $T$; the numerator is not merely a constant times $\cos^2(2\om_{f,k}T)$.

For example, in the short-window limit $T\to0^+$ with fixed scale bounds,
$J_T(\pm\om_{f,k})/J_T(0)\to1$ and
$\mathfrak C_{k,\psi}\to B_k^2/(A_k+B_k)^2$.
For a downward quench $m_i=1$, $m_f=1/2$, $k=0$, this limit is $9/4$.
This explicit case excludes a universal probability bound.

A corresponding Wigner ratio is
\begin{equation}
 \mathfrak C_{k,W}(T;\mathcal A)=
 \frac{\displaystyle\int_{\mathcal A}\dd\Om\,
 \abs{F_{W,k}^{++}(T,\Om)-F_{W,k}^{\rm diag,++}(T,\Om)}^2}
 {\displaystyle\int_{\mathcal A}\dd\Om\,\abs{F_{W,k}^{++}(T,\Om)}^2}.
 \label{eq:wigner-coherence-fraction}
\end{equation}
We choose $\mathcal A=[-6m_i,6m_i]$ in the comparison below and the same
relative-time interval $[-2T,2T]$ for both transforms. These are explicitly
specified, different analysis regions. Their ratios need not agree:
there is no canonical equality between a finite frequency band and a finite
wavelet region. On the full frequency axis Parseval's identity gives the
Wigner ratio directly as the ratio of relative-time squared norms.
An analogous full-scale identity requires an admissible wavelet and the
appropriate admissibility constant, not arbitrary finite regions.

\subsection{What the anomalous wavelet signal measures}

Equation~\eqref{eq:wavelet-coherence-coefficient} exposes the window dependence.
For the uncorrected kernel,
\begin{equation}
 J_\infty(0;b,s)=\sqrt{2\pi}\pi^{-1/4}\sqrt{s}\,
 \ee^{-\omega_0^2/2}.
 \label{eq:morlet-mean}
\end{equation}
Its nonzero mean is small at $\omega_0=6$, but it is not exactly zero.
A strictly zero-mean, admissible Morlet kernel can be defined, up to an
irrelevant common normalization, by
\begin{equation}
 \psi_{\rm adm}(x)=\pi^{-1/4}\ee^{-x^2/2}
 \left(\ee^{\ii\omega_0x}-\ee^{-\omega_0^2/2}\right).
 \label{eq:admissible-morlet}
\end{equation}
Its finite response is
$J_T^{\rm adm}(\om)=J_T^{(\omega_0)}(\om)
-\ee^{-\omega_0^2/2}J_T^{(0)}(\om)$.
For a constant anomalous kernel on the full line,
$J_\infty^{\rm adm}(0)=0$ exactly. At finite $T$, truncation generally gives
$J_T^{\rm adm}(0)\ne0$. The observed anomalous signal is therefore the
response of a finite two-time measurement window to pair coherence,
not a window-independent spectral peak. In the limit $T\to\infty$ at
fixed $b,s$, this response vanishes for the admissible kernel; this limit
does not automatically commute with expanding the observation region.

\begin{figure}[t]
 \centering
 \includegraphics[width=\linewidth]{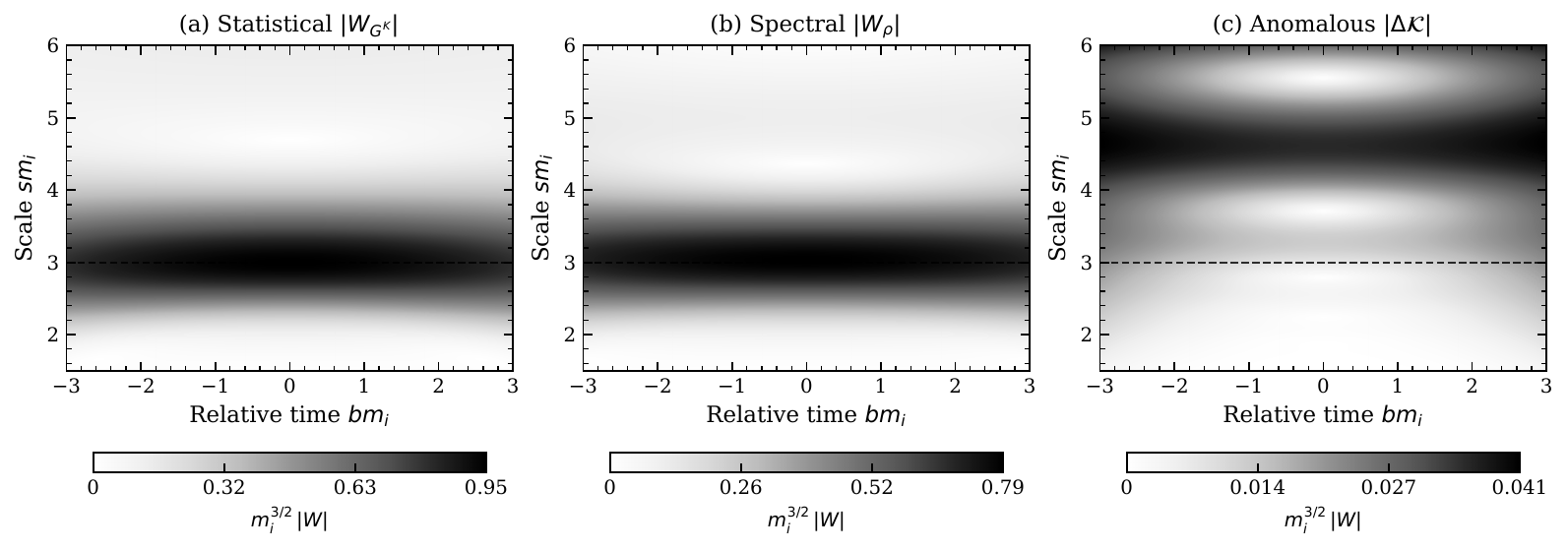}
 \caption{Finite-window Morlet maps at $m_f/m_i=2$, $k=0$, $Tm_i=3$,
 $\omega_0=6$, with relative time restricted to $[-2T,2T]$.
 (a) $|W_{G^K}|=2|\mathcal K|$; (b) $|W_\rho|$; (c) isolated anomalous
 response $|\Delta\mathcal K|$. Each panel has its own grayscale.
 The dashed line is $s_0m_i=3$, not an asserted exact finite-window maximum.
 The spectral kernel is state independent, although its truncated transform
 changes when $T$ changes.}
 \label{fig:keldysh-wavelets}
\end{figure}

Figure~\ref{fig:keldysh-wavelets} displays the actual finite-window statistical,
spectral, and anomalous responses. The third panel is essential: the magnitude
of the full statistical map alone does not isolate anomalous coherence.
The dashed nominal scale identifies the final oscillation, whereas the
anomalous response also contains endpoint-induced scale structure.

\subsection{Exact fixed-mode thermal cancellation}

For an initial Gaussian thermal state with
$N_{i,k}=(\ee^{\beta\om_{i,k}}-1)^{-1}$,
\begin{equation}
 G_k^>(t_1,t_2)=(N_{i,k}+1)u_k(t_1)u_k^*(t_2)
 +N_{i,k}u_k^*(t_1)u_k(t_2).
 \label{eq:thermal-wightman}
\end{equation}
For the same prescribed free evolution at every temperature,
\begin{equation}
 F_k^{(\beta)}=c_k F_k^{(0)},\qquad
 c_k=2N_{i,k}+1=\coth(\beta\om_{i,k}/2),\qquad
 \rho_k^{(\beta)}=\rho_k^{(0)}.
 \label{eq:thermal-factor}
\end{equation}
The superscript $(0)$ here denotes zero temperature. The dephased reference
has the same factor. Linearity and identical windows then imply
\begin{equation}
 \mathcal K^{(\beta)}=c_k\mathcal K^{(0)},\quad
 \Delta\mathcal K^{(\beta)}=c_k\Delta\mathcal K^{(0)},\quad
 \mathfrak C_{k,\psi}^{(\beta)}=\mathfrak C_{k,\psi}^{(0)}.
 \label{eq:thermal-cancellation}
\end{equation}
This is an exact algebraic identity for every finite initial temperature,
with the zero-temperature case obtained by $c_k\to1$. It holds for any
temperature-independent linear smearing, including the Wigner transform.
It also holds for a smooth free quench with a consistently dephased reference.

The hypotheses are restrictive and physically meaningful. The momentum
must be fixed; $c_k$ cannot in general be pulled out of a field-level
momentum integral. The initial covariance must be the common thermal
multiple used above, the evolution must not depend on that temperature,
and the denominator must be nonzero. Self-consistent thermal masses,
different windows, or a general squeezed initial covariance need not obey
the identity. Moreover, a thermally mixed state can have the same ratio
as the vacuum. The ratio diagnoses the shape of anomalous correlations
relative to the chosen reference, not purity, entanglement, or
nonclassical squeezing.

\begin{figure}[t]
 \centering
 \includegraphics[width=\linewidth]{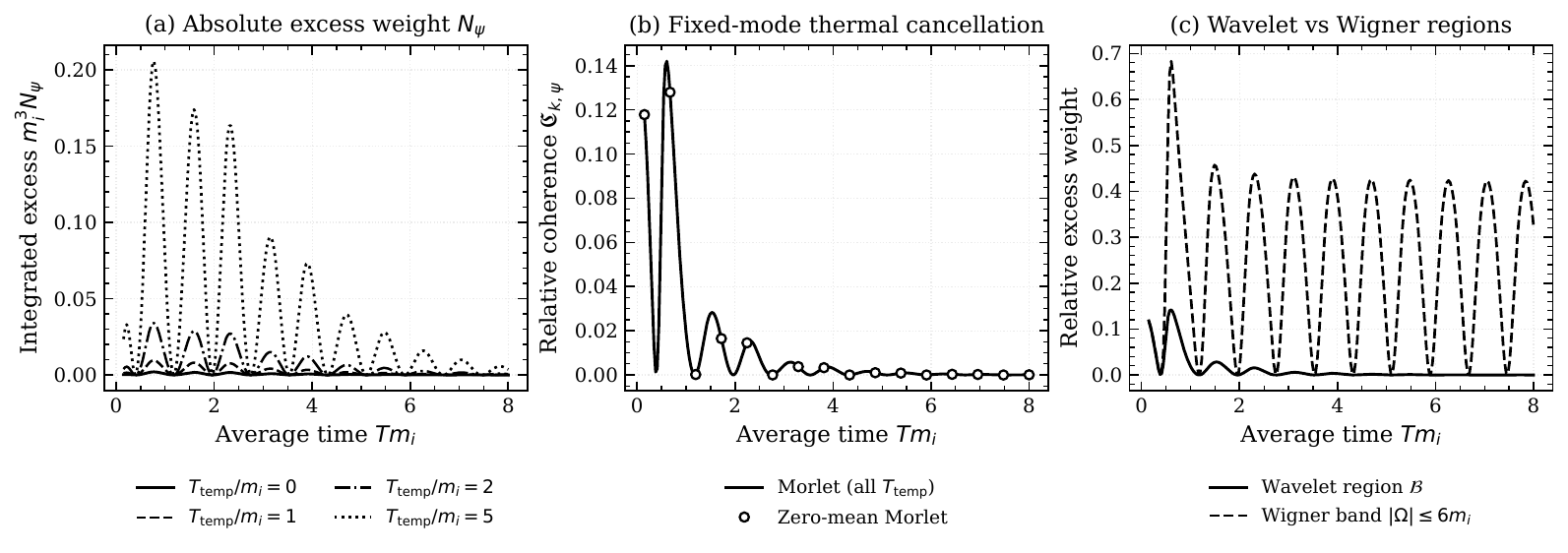}
 \caption{Coherence diagnostics for $m_f/m_i=2$, $k=0$, $\omega_0=6$.
 (a) Integrated excess weight $N_\psi$ at
 $T_{\rm temp}/m_i=0,1,2,5$, using
 $\mathcal B(T)=[-T,T]\times[1.5/m_i,6/m_i]$.
 (b) The normalized curves coincide by Eq.~\eqref{eq:thermal-cancellation};
 circles show the zero-mean Morlet result.
 (c) Comparison with the Wigner ratio on $|\Om|\leq6m_i$.
 Both transforms use $|\tau|\leq2T$, but their finite analysis regions
 differ. The oscillatory zeros follow the anomalous phase, while the
 envelopes and relative magnitudes depend on the observation prescription.}
 \label{fig:thermal-coherence}
\end{figure}

Figure~\ref{fig:thermal-coherence}(a) shows the integrated numerator
$N_\psi=\int_{\mathcal B}\dd b\,\dd s\,|\Delta\mathcal K|^2/s^2$,
rather than an unsmeared anomalous amplitude. Panel (b) illustrates
Eq.~\eqref{eq:thermal-cancellation} and compares the zero-mean kernel.
Panel (c) shows why thermal invariance does not imply identical Wigner
and wavelet ratios: their analysis regions weight the same coherence
differently.

\section{Smooth mass quench}
\label{sec:smooth}

To interpolate between sudden and adiabatic evolution, consider
\begin{equation}
 m^2(t)=m_i^2+\frac{m_f^2-m_i^2}{2}
 \left[1+\tanh\left(\frac{t}{\tau_Q}\right)\right].
 \label{eq:smooth-mass}
\end{equation}
The mode equation remains
\begin{equation}
 \ddot u_k(t)+\left[k^2+m^2(t)\right]u_k(t)=0,
 \label{eq:smooth-mode}
\end{equation}
with the incoming positive-frequency condition in the remote past. The transition time $\tau_Q$ compares the rate of change of the background with the instantaneous oscillation time. Small $\tau_Q$ approaches the sudden problem, whereas large $\tau_Q$ permits adiabatic following.

At late times, the solution can again be projected onto final positive- and negative-frequency modes,
\begin{align}
 \alpha_k&=\ee^{+\ii\om_{f,k}t}
 \frac{\om_{f,k}u_k(t)+\ii\dot u_k(t)}{\sqrt{2\om_{f,k}}},
 \\
 \beta_k&=\ee^{-\ii\om_{f,k}t}
 \frac{\om_{f,k}u_k(t)-\ii\dot u_k(t)}{\sqrt{2\om_{f,k}}}.
 \label{eq:smooth-projection}
\end{align}
The Wronskian implies $\abs{\alpha_k}^2-\abs{\beta_k}^2=1$ independently of $\tau_Q$.

For the hyperbolic-tangent profile, the occupation is known exactly~\cite{Das2015}:
\begin{equation}
 n_k(\tau_Q)=
 \frac{\sinh^2[\pi\tau_Q(\om_{f,k}-\om_{i,k})/2]}
 {\sinh(\pi\tau_Q\om_{i,k})\sinh(\pi\tau_Q\om_{f,k})}.
 \label{eq:smooth-exact}
\end{equation}
Expanding each hyperbolic sine at small $\tau_Q$ recovers
Eq.~\eqref{eq:occupation}. For distinct positive frequencies and large
$\tau_Q$, the leading suppression is
$n_k\sim\exp[-2\pi\tau_Q\min(\om_{i,k},\om_{f,k})]$.
At any fixed $\tau_Q>0$, the ultraviolet tail is therefore different
from the sudden $k^{-4}$ law. The rapid-quench agreement is not uniform
in arbitrarily large momentum.

\begin{figure}[t]
 \centering
 \includegraphics[width=0.90\linewidth]{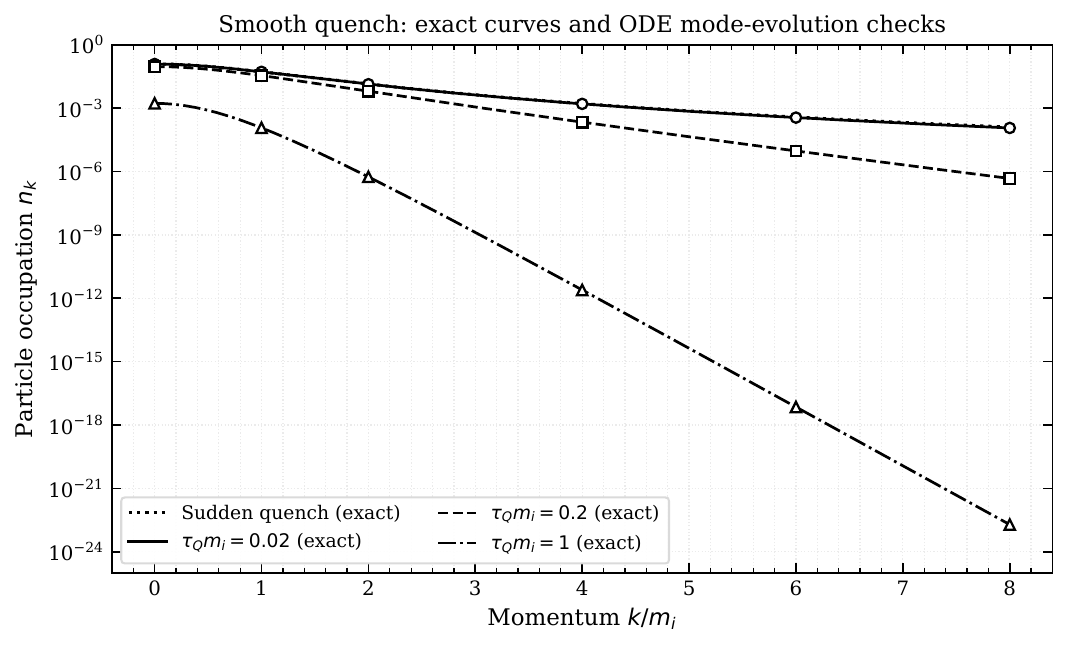}
 \caption{Late-time occupations for $m_f/m_i=2$. Lines are the exact
 sudden and smooth expressions, Eqs.~\eqref{eq:occupation} and
 \eqref{eq:smooth-exact}, at $\tau_Qm_i=0.02,0.2,1$.
 Open markers are independent mode-evolution projections.
 The logarithmic scale displays the adiabatic suppression and the
 nonuniformity of the sudden limit in momentum.}
 \label{fig:occupation-smooth}
\end{figure}

Figure~\ref{fig:occupation-smooth} plots the exact expression and independent
numerical projections from Eq.~\eqref{eq:smooth-projection}. The curves
show suppressed production as the transition is slowed. The numerical
evolution checks the exact solution. The Wronskian is conserved analytically
because the mode equation has a real coefficient; finite-step evolution
approximates that identity.

After a smooth quench has reached its final asymptotic mass, the
Bogoliubov coefficients are generally complex. The vacuum correlator becomes
\begin{equation}
 F_k(T,\tau)=
 \frac{|\alpha_k|^2+|\beta_k|^2}{2\om_{f,k}}\cos(\om_{f,k}\tau)
 +\frac{1}{\om_{f,k}}\Real\!\left[
 \alpha_k\beta_k^*\ee^{-2\ii\om_{f,k}T}\right].
 \label{eq:smooth-anomalous}
\end{equation}
The anomalous amplitude is $\sqrt{n_k(n_k+1)}/\om_{f,k}$.
Its phase also matters: a plot of $n_k$ alone does not determine the
time-resolved coherence. During the transition one must use the
actual product of evolved modes, not Eq.~\eqref{eq:smooth-anomalous}.

\section{From modes to the scalar field}
\label{sec:field}

The field correlator follows by integrating the mode correlators,
\begin{equation}
 F(t_1,t_2;r)=\intk\ee^{\ii kr}F_k(t_1,t_2),
 \qquad
 \rho(t_1,t_2;r)=\intk\ee^{\ii kr}\rho_k(t_1,t_2).
 \label{eq:field-correlators}
\end{equation}
The mode analysis and the momentum integral should be conceptually separated. At fixed $k$, the quench is a finite-dimensional oscillator problem. The integral over $k$ introduces infrared and ultraviolet questions that are absent from the individual mode.

For $m_i,m_f>0$, the infrared sector is regular. The limit $m_i\to0$
requires more than deleting a single continuum momentum point:
the initial field variance behaves as $\int_0\dd k/|k|$ in one spatial
dimension. At finite volume the nonzero modes are regulated by $2\pi/L$,
but the exactly massless zero mode is still a free particle and has no
normalizable oscillator ground state. A small positive initial mass,
or an explicitly prepared zero-mode state together with a finite-volume
prescription, is required. The formulas and plots in this paper use
positive masses. They do not establish an unregulated massless-vacuum quench.

\subsection{Finite volume and the zero mode}

In a periodic interval of length $L$,
\begin{equation}
 k_n=\frac{2\pi n}{L},
 \qquad n\in\mathbb Z,
 \qquad
 F_L(T,\tau)=\frac{1}{L}\sum_{n\in\mathbb Z}F_{k_n}(T,\tau).
 \label{eq:finite-volume}
\end{equation}
The zero mode is one term in this sum. For the sudden transition from $m_i=1$ to $m_f=2$, its occupation is $n_0=0.125$, while its contribution to a number density is $n_0/L$. Thus the zero-mode density decreases as $L^{-1}$ even though the occupation of the individual oscillator remains fixed. This is why the zero mode should not be treated as an ordinary point in a continuum momentum integral.

\section{Ultraviolet structure in \texorpdfstring{$1+1$}{1+1} dimensions}
\label{sec:uv}

The excitation energy relative to the final vacuum is, in $1+1$ dimensions,
\begin{equation}
 \Delta\mathcal E_\Lambda=\frac{1}{\pi}\int_0^\Lambda\dd k\,
 \om_{f,k}n_k.
 \label{eq:excitation-energy}
\end{equation}
Using Eq.~\eqref{eq:high-k-beta}, the large-$k$ integrand behaves as
\begin{equation}
 \frac{1}{\pi}\om_{f,k}n_k
 =\frac{(m_f^2-m_i^2)^2}{16\pi k^3}+\cO(k^{-5}).
 \label{eq:uv-1d-integrand}
\end{equation}
Consequently the cutoff limit is finite and the residual tail satisfies
\begin{equation}
 \Delta\mathcal E_\infty-\Delta\mathcal E_\Lambda
 =\frac{(m_f^2-m_i^2)^2}{32\pi\Lambda^2}+\cO(\Lambda^{-4}).
 \label{eq:uv-1d-tail}
\end{equation}
For $m_i=1$ and $m_f=2$, the coefficient of the asymptotic tail is $9/(32\pi)$. The logarithmic sensitivity encountered in $3+1$ dimensions is therefore absent from the present $1+1$-dimensional particle-production contribution.

This is a deliberately restricted statement. It concerns the particle-production contribution relative to the final vacuum. It is not a complete renormalization of $\avg{T_{\mu\nu}}$, which still requires the vacuum terms, local counterterms, and a consistent renormalization prescription for the full stress tensor. The corresponding $3+1$-dimensional ultraviolet logarithm is outside the scope of this article.

\section{Weak interactions: a Hartree extension}
\label{sec:hartree}

Consider a vanishing mean field with interaction
\begin{equation}
 \mathcal L=\frac12(\partial_t\phi)^2-\frac12(\partial_x\phi)^2
 -\frac12m_R^2(t)\phi^2-\frac{\lambda}{4!}:\!\phi^4\!:_i.
 \label{eq:lambda4}
\end{equation}
The normal-ordering reference is the initial vacuum of mass $m_i>0$.
This fixes the finite mass convention as well as removing the tadpole
divergence. At Hartree level~\cite{CJT,Berges,Cooper:1994hr},
\begin{align}
 M_H^2(t)&=m_R^2(t)+\frac{\lambda}{2}\Delta F_H(t),\nonumber\\
 \Delta F_H(t)&=\intk\left[|u_{H,k}(t)|^2-\frac{1}{2\om_{i,k}}\right],
 \label{eq:hartree-mass}\\
 \ddot u_{H,k}+[k^2+M_H^2(t)]u_{H,k}&=0.
 \label{eq:hartree-mode}
\end{align}
These equations are self-consistent. A regulator is understood before
subtraction; for the vacuum initial condition in $1+1$ dimensions,
the subtraction removes the logarithmic tadpole. At finite volume,
$\intk$ is replaced by $L^{-1}\sum_k$ with the same reference.

When $m_R^2$ is constant, the corresponding Gaussian energy density,
up to a time-independent vacuum subtraction, is
\begin{equation}
 \mathcal E_H=
 \frac12\intk\left[|\dot u_{H,k}|^2+
 (k^2+m_R^2)|u_{H,k}|^2\right]
 +\frac{\lambda}{8}(\Delta F_H)^2.
 \label{eq:hartree-energy}
\end{equation}
Differentiating at fixed regulator and using Eq.~\eqref{eq:hartree-mode}
shows that the interaction term cancels the mass-feedback contribution,
so $\dot{\mathcal E}_H=0$ after the quench.
A stationary vacuum would instead satisfy the subtracted gap equation
\begin{equation}
 M^2=m_R^2+\frac{\lambda}{2}\intk
 \left[\frac{1}{2\sqrt{k^2+M^2}}-\frac{1}{2\om_{i,k}}\right].
 \label{eq:gap}
\end{equation}
The un-subtracted gap equation and a subtracted dynamical mass must not
be combined without changing the mass convention.

\subsection{Leading correction around the free quench}

Replacing $u_H$ by the free solution on the right-hand side gives
a first-order mass correction, not an exact self-consistent solution:
\begin{equation}
 \delta M_{(1)}^2(t)=\frac{\lambda}{2}\intk
 B_k\left[\cos(2\om_{f,k}t)-1\right].
 \label{eq:hartree-first}
\end{equation}
The integrand is ultraviolet integrable because $B_k=\cO(|k|^{-3})$.
For a mass increase and $\lambda>0$, $B_k>0$, so the time-averaged
shift in this initial-vacuum subtraction scheme is negative,
\begin{equation}
 \overline{\delta M_{(1)}^2}=-\frac{\lambda}{2}\intk B_k<0.
 \label{eq:hartree-mean}
\end{equation}
Within a weak-shift oscillatory approximation,
$\overline\Omega_k^2=\om_{f,k}^2+\overline{\delta M_{(1)}^2}$.
The corresponding nominal Morlet scale increases, rather than contracts.
This comparison is scheme-specific: it is relative to $m_R^2=m_f^2$
in Eq.~\eqref{eq:lambda4}, not to an un-subtracted bare mass.

A first-order mode correction that does not assume slow modulation is
\begin{equation}
 \delta u_k(t)=
 -\int_0^t\dd t'\,
 \frac{\sin[\om_{f,k}(t-t')]}{\om_{f,k}}\,
 \delta M_{(1)}^2(t')u_k^{(0)}(t').
 \label{eq:hartree-retarded}
\end{equation}
It has zero initial correction and follows by applying the free retarded
Green function to the perturbed mode equation. It is valid while the
perturbative correction remains small; secular terms at late times
require resummation or self-consistent evolution.
Substitution into $u(t_1)u^*(t_2)$ then determines the correction to
the bilocal statistical response.

The mass feedback contains all internal frequencies $2\om_{f,q}$.
One cannot replace this spectrum by $2\om_{f,k}$ or assert universal
sidebands at $\pm2\om_{f,k}$ in relative-time frequency. Average-time
modulation and relative-time frequency shifts are different statements.
Their relation must be obtained from Eq.~\eqref{eq:hartree-retarded}
and the chosen window.

\subsection{What Hartree does not describe}

The double-bubble term is two-loop in the 2PI effective action and yields
a one-loop, local, real tadpole self-energy. It has no nonlocal collision
kernel and hence no collisional width. This does not forbid dephasing,
time-dependent mode amplitudes, or decay of a coarse-grained observable.
For $\lambda\phi^4$ theory, the nonlocal setting-sun self-energy is
order $\lambda^2$ and two-loop as a self-energy; it comes from the
three-loop 2PI basketball diagram~\cite{Berges,Berges:2001fi}.
Including it allows scattering, but does not by itself prove complete
thermalization for every state or truncation. The present section
establishes a consistent extension and its leading correction, not
an interacting thermalization result or a computed Hartree wavelet spectrum.

\section{A fixed expanding background}
\label{sec:frw}

As a separate extension, consider a prescribed $1+1$-dimensional
de Sitter geometry~\cite{BirrellDavies,MukhanovWinitzki},
\begin{equation}
 \dd s^2=-\dd t^2+a^2(t)\dd x^2,\qquad
 a(t)=\ee^{H(t-t_0)},\qquad R=2H^2.
 \label{eq:frw-metric}
\end{equation}
For $\mu^2=m^2+\xi R\geq0$,
\begin{equation}
 \ddot u_k+H\dot u_k+
 \left[\frac{k^2}{a^2}+\mu^2\right]u_k=0,\qquad
 a(t)(u_k\dot u_k^*-\dot u_k u_k^*)=\ii.
 \label{eq:frw-mode}
\end{equation}
The Hubble term is geometric dilution, not a phenomenological dissipative
coupling. The canonical variable $\chi_k=\sqrt{a}\,u_k$ satisfies
\begin{equation}
 \ddot\chi_k+\Omega_{\chi,k}^2(t)\chi_k=0,\qquad
 \Omega_{\chi,k}^2=\frac{k^2}{a^2}+\mu^2-\frac{H^2}{4}.
 \label{eq:chi-mode}
\end{equation}
The $-H^2/4$ term is essential for the late-time classification.

\subsection{Exact modes and state selection}

In conformal time $\eta=-[Ha(t)]^{-1}$, the mode equation becomes
$u_k''+[k^2+\mu^2/(H^2\eta^2)]u_k=0$.
For $k\ne0$, the early-time positive-frequency solution is
\begin{equation}
 u_k(\eta)=\frac{\sqrt{-\pi\eta}}{2}
 \ee^{\ii\pi(2\nu+1)/4}H_\nu^{(1)}(-|k|\eta),
 \qquad \nu=\sqrt{\frac14-\frac{\mu^2}{H^2}}.
 \label{eq:hankel}
\end{equation}
The branch has $\Real\nu\geq0$ and $\Imag\nu\geq0$.
The prefactor also normalizes imaginary order: its modulus is
$\ee^{-\pi\Imag\nu/2}$. At $-|k|\eta\gg1$,
$u_k\to\ee^{-\ii|k|\eta}/\sqrt{2|k|}$, which fixes the canonical Wronskian.
This is an explicit state prescription. Other Gaussian states require
their own mode combination.

Alternatively, at a finite initial time where $\Omega_\chi^2>0$
and the adiabatic conditions hold, a leading WKB prescription is
\begin{equation}
 u_k(t_0)=\frac{1}{\sqrt{2a(t_0)\Omega_{\chi,k}(t_0)}},\qquad
 \dot u_k(t_0)=
 \left[-\frac H2-\frac{\dot\Omega_{\chi,k}}{2\Omega_{\chi,k}}
 -\ii\Omega_{\chi,k}\right]_{t_0}u_k(t_0).
 \label{eq:frw-initial}
\end{equation}
The amplitude derivative is required by that WKB ansatz.
This finite-time prescription need not coincide exactly with
Eq.~\eqref{eq:hankel}.

\subsection{Scale drift and the limits of a local-frequency picture}

In an oscillatory regime with negligible variation across the analysis
window, a component of relative-time phase has nominal scale
\begin{equation}
 s_0(T)\simeq\frac{\omega_0}{\Omega_{\chi,k}(T)}.
 \label{eq:desitter-scale-drift}
\end{equation}
This is a local approximation, not an identity for the exact wavelet
ridge. Sufficient conditions include $Hs\ll1$,
$|\dot\Omega_\chi|s/\Omega_\chi\ll1$, and a window well separated
from its endpoints. Deep inside the horizon means $|k|/(aH)\gg1$;
it does not mean $H(T-t_0)\ll1$.
For a single locally monochromatic component the normalization shift
in Eq.~\eqref{eq:exact-scale} must also be included when locating
the magnitude maximum.

The exact late-time solutions of Eq.~\eqref{eq:frw-mode} distinguish
three regimes:
\begin{equation}
 u_k(t)\ \sim\
 \begin{cases}
 a^{-1/2}\big(C_+\ee^{+\ii\varpi t}+C_-\ee^{-\ii\varpi t}\big),
 &\mu^2>H^2/4,\quad \varpi=\sqrt{\mu^2-H^2/4},\\
 a^{-1/2}(C_1+C_2t),&\mu^2=H^2/4,\\
 C_+a^{-1/2+\nu}+C_-a^{-1/2-\nu},
 &0\leq\mu^2<H^2/4.
 \end{cases}
 \label{eq:desitter-late}
\end{equation}
Heavy modes remain oscillatory, with frequency $\varpi$, not $\mu$.
A late finite scale near $\omega_0/\varpi$ is meaningful only when the
local oscillatory conditions hold, for example for $\varpi\gg\omega_0H$.
Light modes are nonoscillatory in this limit, so there is no universal
finite-frequency Morlet ridge that freezes at $\omega_0/\mu$.
At the threshold the independent solutions contain a logarithm of $a$.

In two spacetime dimensions the massless conformal case is $m=0,\xi=0$.
For each nonzero $k$,
$u_k=\ee^{-\ii|k|\eta}/\sqrt{2|k|}$ exactly.
Its cosmic-time instantaneous frequency $|k|/a$ tends to zero, while
the mode approaches a constant. The subhorizon estimate
$s_0\sim\omega_0 a/|k|$ grows until the local-frequency approximation fails;
it does not approach a finite universal scale.
Mode freezing, when present, must not be identified with automatic
classicality of the quantum state.

These results concern a free field on a fixed geometry.
Interacting infrared dynamics and stochastic descriptions in de Sitter
space~\cite{Starobinsky:1994bd,Garbrecht:2014dca} address additional
questions and do not establish the wavelet-ridge claims above.
Taking $H\to0$ in the mode equation and its canonical normalization
recovers the Minkowski oscillator. No stress-tensor backreaction
or semiclassical Einstein equation is solved here.

\section{Numerical checks and the smooth-quench benchmark}
\label{sec:results}

Table~\ref{tab:results} gives zero-momentum occupations evaluated from
the exact expressions for $m_i=1$, $m_f=2$. It quantifies the dynamical
suppression illustrated by Fig.~\ref{fig:occupation-smooth}, independently
of any wavelet convention.

\begin{table}[t]
 \centering
 \caption{Zero-momentum occupation for $m_f/m_i=2$. The sudden entry is
 exact; the displayed smooth values are rounded evaluations of the exact
 expression in Eq.~\eqref{eq:smooth-exact}, not fitted parameters.}
 \label{tab:results}
 \begin{tabular}{lll}
 \toprule
 $\tau_Qm_i$ & $n_0$ & Physical regime\\
 \midrule
 $0$ & $1/8$ & Sudden quench\\
 $0.02$ & $0.1246$ & Rapid smooth quench\\
 $0.2$ & $0.09421$ & Reduced production\\
 $1$ & $1.713\times10^{-3}$ & Strongly suppressed production\\
 $5$ & $2.271\times10^{-14}$ & Adiabatically suppressed production\\
 \bottomrule
 \end{tabular}
\end{table}

The open markers in Fig.~\ref{fig:occupation-smooth} are obtained by independent
numerical integration of the mode equation~\eqref{eq:smooth-mode} from the in-vacuum
state in the remote past to asymptotic late times, followed by projection onto the
out-basis via Eq.~\eqref{eq:smooth-projection}. Across all sampled momenta and
switching durations, the numerical solutions agree with the exact analytic result
in Eq.~\eqref{eq:smooth-exact} to high precision, verifying both Wronskian
conservation and the analytical Bogoliubov identity. The plotted curves display
the exact analytical expressions.

The wavelet maps and coherence ratios in Figs.~\ref{fig:keldysh-wavelets} and
\ref{fig:thermal-coherence} are evaluated using the exact closed-form finite-window
kernel in Eq.~\eqref{eq:J-finite}. Numerical quadrature of the integral definition
in Eq.~\eqref{eq:J-definition} confirms the closed-form expression across all
tested parameter regions, with discretization uncertainties negligible on the scale
of the figures.

\section{Discussion}
\label{sec:discussion}

The quench produces occupations and anomalous correlations.
The former are time independent in the final free-particle basis;
the latter retain an oscillatory average-time phase.
At fixed momentum, neither a sudden quench nor the late free evolution
provides collisional equilibration. This does not exclude dephasing
after integration over momentum: local observables can approach
a stationary generalized description while individual modes retain
coherence. The dephased reference is therefore not automatically
a Gibbs equilibrium state.

The bilocal observable in Eq.~\eqref{eq:wavelet-operator} specifies what
a relative-time and scale analysis measures. Figures~\ref{fig:keldysh-wavelets}
and \ref{fig:thermal-coherence} separate the full statistical response
from its excess over the same-window dephased reference.
The decreasing wavelet envelope in Fig.~\ref{fig:thermal-coherence}(c)
is not evidence of dynamical decoherence: $B_k$ remains constant and
the underlying anomalous oscillation persists. The changing overlap
with the observation window accounts for this decrease.
The observed anomalous structure combines physical pair coherence with
the transfer function of the observation window. Both must be given
to make results from different analyses comparable. In particular, a
zero-mean wavelet on an unbounded relative-time domain is blind to the
constant anomalous term of the sudden-quench correlator.

The thermal cancellation is useful precisely because its scope is clear.
For a fixed mode in free evolution, a normalized quadratic functional
of a linearly smeared thermal correlator removes the common factor
$2N_{i,k}+1$. The absolute signal still grows with temperature.
Thus the cancellation neither isolates a vacuum state nor demonstrates
entanglement; it isolates a relative shape property under a specified
family of initial states. A momentum-integrated observable or an
interacting self-consistent evolution requires a new analysis.

The Wigner comparison gives two further lessons. First, negative
interference lobes are not an inconsistency to be repaired by another
transform. Second, positivity of the Gaussian extension in
Fig.~\ref{fig:wigner-gaussian} is a property of that extension and its
width, not a positivity statement about the original two-time measurement.
Finite-band Wigner and finite-region wavelet ratios generally differ
even when both use exactly the same relative-time data.

\subsection{Localization, resolution, and comparison with Wigner analysis}
\label{sec:wavelet-advantages}

The wavelet representation provides a family of localized correlation
measurements with scale-dependent relative-frequency resolution. Its utility
must be distinguished from positivity of a representation or removal of an
ultraviolet divergence. The nonnegative quantity
$|\mathcal K_{k,\psi}(T;b,s)|^2$ is a squared correlation response. It is not
the field's Hamiltonian energy density. In the present normalization
$[\mathcal K]=({\rm mass})^{-3/2}$, whereas a physical energy density in
$1+1$ dimensions has dimension $({\rm mass})^2$. Taking a modulus also
discards phase information, and the analogous squared modulus of a Fourier
response is nonnegative. Thus positivity of this map does not establish a
wavelet-specific resolution of the signed Wigner spectrum or of ultraviolet
renormalization.

Localization gives a quantitative control of truncation for bounded input.
Let $F_{\rm ext}$ denote a specified extension with
$|F_{\rm ext}(T,\tau)|\le M$, and compare its full-line uncorrected Morlet
coefficient with the coefficient restricted to $[-2T,2T]$. For
$d_\pm=2T\mp b>0$, the Gaussian envelope gives
\begin{align}
 |W_\infty-W_{[-2T,2T]}|
 \le M\pi^{-1/4}\sqrt{\frac{\pi s}{2}}
 \left[\operatorname{erfc}\left(\frac{d_+}{\sqrt2s}\right)
       +\operatorname{erfc}\left(\frac{d_-}{\sqrt2s}\right)\right].
\end{align}
This is an absolute error bound, not a bound on relative error near a zero
of the response. For $x>0$,
\[
 \int_x^\infty \ee^{-u^2}\dd u
 \leq \frac{1}{x}\int_x^\infty u\ee^{-u^2}\dd u
 =\frac{\ee^{-x^2}}{2x}.
\]
The inequality follows pointwise from $u/x\geq1$. It shows explicitly
why the bound becomes exponentially small for $d_\pm/s\gg1$.
A distance of two scale widths alone need not make the error negligible:
at $b=0$ and $2T/s=2$, the omitted fraction of the Gaussian envelope's
integral is $\operatorname{erfc}(\sqrt2)\simeq0.04550$.
This number evaluates an envelope bound, not a measured noise level.
A Gaussian-window Fourier analysis also admits localization bounds.
Indeed, if
$S_s(b,\Omega)=\int_{-2T}^{2T}\dd\tau\,
 F(T,\tau)\exp[-(\tau-b)^2/(2s^2)]\exp(-\ii\Omega\tau)$,
then the uncorrected Morlet coefficient obeys exactly
\begin{equation}
 W(T;b,s)=\pi^{-1/4}s^{-1/2}\ee^{\ii\omega_0 b/s}
 S_s(b,\omega_0/s).
\end{equation}
The distinction from a fixed-width Fourier analysis is the prescribed
coupling of frequency and window width, not exclusive access to the data.
Moreover, suppressing endpoints suppresses part of the anomalous signal
studied here, since its admissible-wavelet response on the full line
vanishes exactly.

The crossed-sector correlator in Eq.~\eqref{eq:crossed} contains the
hybrid frequencies $\Omega_\pm=(\om_{f,k}\pm\om_{i,k})/2$. Its Fourier
transform exists as a tempered distribution; Abel damping is one useful
representation of that limit. A transform restricted to a finite interval
does not require this limiting prescription. A crossed-sector wavelet
analysis would instead require an observable extending beyond the
post-quench interval used in Eq.~\eqref{eq:wavelet-operator}.
For example, the positive crossed-sector response is
\begin{equation}
 W^{+-}(T;b,s)=\int_{2T}^{\infty}\dd\tau\,
 \psi_{b,s}^*(\tau)F_{k,+-}(T,\tau).
\end{equation}
It is finite at each finite scale because of Gaussian localization, but
its amplitudes include the endpoint response and overlap between the
two frequency components. Extracting $\alpha_k$ and $\beta_k$ requires
accounting for those factors and specifying a resolution criterion.

Even an isolated full-line positive-frequency exponential has a coefficient
proportional to $\sqrt{s}\exp[-(s\Omega-\omega_0)^2/2]$. For $\Omega>0$,
its maximum occurs at
\begin{equation}
 s_* = \frac{\omega_0+\sqrt{\omega_0^2+2}}{2\Omega},
\end{equation}
rather than exactly at the nominal scale $\omega_0/\Omega$.
Real oscillations and a sector boundary introduce further response terms.
Appendix~\ref{app:cross-wavelet} gives the exact crossed-sector kernel and
bounds its difference from the full-line response.
For $m_f>m_i>0$, the ratio of the unwanted positive-frequency
$\beta_k$ contribution to the $\alpha_k$ contribution at the nominal
scale $s_+=\omega_0/\Omega_+$, in the full-line extension, is
\begin{equation}
 r_{+\leftarrow-}=
 \left|\frac{\beta_k}{\alpha_k}\right|
 \exp\left[-\frac{\omega_0^2}{2}
       \left(1-\frac{\Omega_-}{\Omega_+}\right)^2\right].
 \label{eq:cross-leakage}
\end{equation}
At $k=0$, $m_f/m_i=100$, and $\omega_0=6$, this is exactly
$(99/101)\exp(-72/10201)$. The inequality $\ee^{-x}\geq1-x$
gives the rational lower bound
\[
 r_{+\leftarrow-}\geq
 \frac{99}{101}\left(1-\frac{72}{10201}\right)>\frac{97}{100}.
\]
Here $\ee^{-x}\geq1-x$ follows by integrating
$1-\ee^{-v}\geq0$ over $0\leq v\leq x$.
The numerical value $r_{+\leftarrow-}\simeq0.9733$ evaluates this exact
expression. The large overlap persists for a crossed-sector window
sufficiently far from its endpoint, as quantified in the appendix.
Thus two peak heights cannot universally be identified with independent
Bogoliubov amplitudes. This statement concerns component mixing,
not a theorem about the number of maxima of the full response.

Finally, the affine integration measure is $\dd b\,\dd s/s^2$.
It is dimensionless and invariant under simultaneous dilations of $b$ and
$s$. The logarithmic scale coordinate $x=\ln(s/s_{\rm ref})$ satisfies
$\dd s/s=\dd x$, whereas
$\dd s/s^2=\ee^{-x}\dd x/s_{\rm ref}$.
Logarithmic sampling can conveniently track redshift over several scales,
within the adiabatic conditions stated in the de Sitter section. It does
not by itself implement a renormalization-group transformation.

These properties motivate a controlled comparison using the same temporal
data and an explicit resolution criterion. The figures in this work
compare the stated finite analysis regions; they do not establish a
universal operational superiority of one representation.

Finally, the extensions delimit rather than enlarge the exact claim.
Hartree adds a collective, state-dependent mass and no collisional
self-energy. Its first-order shift depends on the specified subtraction
scheme. De Sitter redshift permits a local scale interpretation only
in an oscillatory regime; the exact Hankel solutions exclude a universal
finite-scale freezing rule. Computing self-consistent interacting
maps and testing their resolution dependence would be a distinct next
step, not a consequence already established by the free-quench figures.

\section{Conclusions}
\label{sec:conclusions}

The main result is an explicitly defined finite-window Keldysh observable
whose dephased excess resolves anomalous pair coherence in relative time
and scale. Its fixed-mode thermal cancellation is exact for a common
Gaussian thermal amplitude and temperature-independent free evolution.
Its dependence on the observation window is equally essential:
a full-line admissible wavelet removes the constant anomalous term.

The exact finite-window kernel makes these statements quantitative
and supports a comparison with Wigner analysis without assigning either
representation a probability interpretation. Smooth-quench occupations
supply an independent exact reference for the suppression of pair production.
Gaussian localization bounds the absolute endpoint error, but does not
guarantee separate spectral components. The exact crossed-sector response
mixes both Bogoliubov amplitudes; Eq.~\eqref{eq:cross-resolution-condition}
states a sufficient condition for approximate isolation.
Positive initial masses keep the examples infrared regular; the finite
excitation-energy tail in $1+1$ dimensions does not replace the
renormalization of general local operators.

The Hartree and de Sitter sections provide equations and validity
conditions for extending this construction, not evidence of complete
thermalization or universal scale freezing. The resulting framework
offers a reproducible baseline for asking which multiscale signatures
survive changes of window, initial state, and dynamics.

\appendix

\section{Bogoliubov matching and the Wronskian}
\label{app:matching}

At $t=0^-$,
\begin{equation}
 u_k(0^-)=\frac{1}{\sqrt{2\om_{i,k}}},
 \qquad
 \dot u_k(0^-)=-\ii\sqrt{\frac{\om_{i,k}}{2}}.
 \label{eq:appendix-initial}
\end{equation}
At $t=0^+$, Eq.~\eqref{eq:post-mode} gives
\begin{equation}
 u_k(0^+)=\frac{\alpha_k+\beta_k}{\sqrt{2\om_{f,k}}},
 \qquad
 \dot u_k(0^+)=-\ii\sqrt{\frac{\om_{f,k}}{2}}(\alpha_k-\beta_k).
 \label{eq:appendix-final}
\end{equation}
Equating these expressions yields Eq.~\eqref{eq:matching}. The Wronskian of the final mode is
\begin{equation}
 u_k\dot u_k^*-\dot u_k u_k^*=\ii\left(\abs{\alpha_k}^2-\abs{\beta_k}^2\right),
 \label{eq:appendix-wronskian}
\end{equation}
which proves Eq.~\eqref{eq:unitarity}.

\section{Ultraviolet tail in \texorpdfstring{$1+1$}{1+1} dimensions}
\label{app:uv}

Using Eq.~\eqref{eq:high-k-beta},
\begin{equation}
 \om_{f,k}n_k=\frac{(m_f^2-m_i^2)^2}{16k^3}+\cO(k^{-5}).
 \label{eq:appendix-energy-asymptotic}
\end{equation}
In one spatial dimension, the two sides of the momentum axis give
\begin{align}
 \Delta\mathcal E_\Lambda
  &\sim \frac{1}{\pi}\int^\Lambda\dd k
  \frac{(m_f^2-m_i^2)^2}{16k^3}
  =\frac{(m_f^2-m_i^2)^2}{32\pi}
  \left(\frac{1}{k_{\mathrm{IR}}^2}-\frac{1}{\Lambda^2}\right),
 \end{align}
where $k_{\mathrm{IR}}$ denotes the lower limit of the ultraviolet asymptotic region. The $\Lambda^{-2}$ approach to the finite limit is the content of Eq.~\eqref{eq:uv-1d-tail}; no logarithmic subtraction is required for this contribution in $1+1$ dimensions.

\section{Sign and time average of the leading Hartree insertion}
\label{app:hartree-sign}

The post-quench mode can also be written as
\begin{equation}
 u_k^{(0)}(t)=\frac{1}{\sqrt{2\om_{i,k}}}
 \left[\cos(\om_{f,k}t)-\ii\frac{\om_{i,k}}{\om_{f,k}}
 \sin(\om_{f,k}t)\right].
\end{equation}
Taking its squared modulus gives the exact identity
\begin{equation}
 |u_k^{(0)}(t)|^2-\frac{1}{2\om_{i,k}}
 =-2B_k\sin^2(\om_{f,k}t).
\end{equation}
For $m_f>m_i>0$, $B_k$ is positive. The integral of this expression
is strictly negative at every $t>0$, not just nonpositive.
To see this without time sampling, choose an integer $n$ with
$n\pi+\pi/4>m_ft$ and define positive endpoints
\begin{equation}
 q_-^2=\left(\frac{n\pi+\pi/4}{t}\right)^2-m_f^2,
 \qquad
 q_+^2=\left(\frac{n\pi+3\pi/4}{t}\right)^2-m_f^2.
\end{equation}
For $q\in[q_-,q_+]$, $\sin^2(\om_{f,q}t)\geq1/2$.
Since $B_q$ decreases for $q>0$, symmetry and the nonpositive
integrand imply
\begin{equation}
 \Delta F^{(0)}(t)
 =-\frac{2}{\pi}\int_0^\infty\dd q\,B_q\sin^2(\om_{f,q}t)
 \leq-\frac{B_{q_+}}{\pi}(q_+-q_-)<0.
\end{equation}
This explicit interval supplies the positive-measure support needed
for strict negativity. The origin is separate: $\Delta F^{(0)}(0)=0$.

For an averaging time $\mathcal T>0$, absolute integrability permits
interchanging the time and momentum integrals:
\begin{equation}
 \frac{1}{\mathcal T}\int_0^{\mathcal T}\dd t\,\Delta F^{(0)}(t)
 =-\intk B_k+\intk B_k
 \frac{\sin(2\om_{f,k}\mathcal T)}{2\om_{f,k}\mathcal T}.
\end{equation}
The average is strictly negative because its time integrand is
strictly negative for $t>0$. Furthermore,
\begin{equation}
 \left|\frac{1}{\mathcal T}\int_0^{\mathcal T}\dd t\,
 \Delta F^{(0)}(t)+\intk B_k\right|
 \leq\frac{1}{2\mathcal T}\intk\frac{B_k}{\om_{f,k}}.
\end{equation}
The right-hand integral is finite, proving the stated infinite-time
mean and Eq.~\eqref{eq:hartree-mean}. Multiplication by $\lambda/2>0$
preserves the sign of this first-order insertion. Neither statement
assumes or proves a sign for a fully self-consistent Hartree solution.

For completeness, $B_q\leq(m_f^2-m_i^2)/(4q^3)$ for $q>0$ gives
the explicit tail bounds
\begin{equation}
 \int_{|q|>\Lambda}\frac{\dd q}{2\pi}B_q
 \leq\frac{m_f^2-m_i^2}{8\pi\Lambda^2},\qquad
 |\Delta F^{(0)}_{\rm tail}(t)|
 \leq\frac{m_f^2-m_i^2}{4\pi\Lambda^2}.
\end{equation}
They bound truncation errors, while finite-interval quadrature still
requires its own error estimate.

\section{Physical scope of the approximations}
\label{app:scope}

The hierarchy of statements in this article is:
\begin{enumerate}[label=\arabic*.]
 \item The sudden free-field mode solution and its correlators are exact.
 \item The smooth free-field mode equation is exact, and its transition amplitudes are defined by late-time matching to the final positive- and negative-frequency basis.
 \item The Wigner and wavelet maps are representations of the same two-time correlator; in particular, the wavelet map is directly a multiscale representation of the Keldysh statistical correlator.
 \item The finite-volume and ultraviolet discussions specify the infrared and cutoff structure but do not automatically renormalize local composite operators.
 \item Hartree theory is a Gaussian truncation of the interacting theory.
 \item The FRW calculation is quantum field theory on a prescribed background, not a solution of semiclassical gravity.
\end{enumerate}
Keeping these statements separate prevents a successful mode calculation from being mistaken for a complete solution of the interacting, renormalized, gravitationally backreacting problem.

\section{Crossed-sector response and a controlled endpoint limit}
\label{app:cross-wavelet}

This appendix uses the uncorrected Morlet kernel and the vacuum correlator.
A common initial thermal factor multiplies every term and does not change
the component ratios. Set $L=2T$, with $T\geq0$, and define
\begin{align}
 \mathcal G_L(q;b,s)
 &=\int_L^\infty\dd\tau\,\psi_{b,s}^*(\tau)\ee^{\ii q\tau}
 \nonumber\\
 &=\pi^{1/4}\sqrt{\frac{s}{2}}\,
 \ee^{\ii qb-(sq-\omega_0)^2/2}
 \operatorname{erfc}\left(
 \frac{L-b}{\sqrt2s}-\frac{\ii(sq-\omega_0)}{\sqrt2}\right).
 \label{eq:cross-G}
\end{align}
Completing the Gaussian square gives Eq.~\eqref{eq:cross-G}.
Its derivative with respect to $L$ is
$-\psi_{b,s}^*(L)\ee^{\ii qL}$.
The boundary condition at $L\to+\infty$ must also be retained.
For fixed real $Y$ and $X>0$, analyticity gives
\[
 |\operatorname{erfc}(X+\ii Y)|
 \leq\operatorname{erfc}(X)
 +\frac{2|Y|}{\sqrt\pi}\ee^{-X^2+Y^2}\longrightarrow0.
\]
The second term bounds the integral along the vertical segment from
$X$ to $X+\ii Y$. Together with the derivative, this boundary condition
fixes the half-line integral uniquely.

Using Eq.~\eqref{eq:crossed} and expanding each cosine, one obtains
\begin{align}
 W^{+-}
 =\frac{1}{4\sqrt{\om_{f,k}\om_{i,k}}}\big[
 &\alpha_k\{\ee^{2\ii\Omega_-T}\mathcal G_L(\Omega_+)
          +\ee^{-2\ii\Omega_-T}\mathcal G_L(-\Omega_+)\}
 \nonumber\\
 {}+&\beta_k\{\ee^{2\ii\Omega_+T}\mathcal G_L(\Omega_-)
          +\ee^{-2\ii\Omega_+T}\mathcal G_L(-\Omega_-)\}\big].
 \label{eq:cross-W-exact}
\end{align}
The suppressed arguments $(b,s)$ are identical in all four terms.
The endpoint and both frequency signs are therefore part of the
measurement, even when only one nominal scale is examined.

For $b>L$, let $d=b-L$, $C_s=\sqrt2\pi^{1/4}\sqrt{s}$,
$h_q=\exp[-(sq-\omega_0)^2/2]$, and
$\delta_d=\tfrac12\operatorname{erfc}(d/(\sqrt2s))$.
The omitted Gaussian tail gives
\begin{equation}
 |\mathcal G_L(q)-\mathcal G_{-\infty}(q)|
 \leq C_s\delta_d,\qquad
 \mathcal G_{-\infty}(q)=C_s\ee^{\ii qb}h_q.
 \label{eq:cross-endpoint-bound}
\end{equation}
This is an absolute, frequency-independent error bound.
If $h_{\Omega_+}>\delta_d$, the positive-frequency mixing ratio obeys
\begin{align}
 \left|\frac{\beta_k}{\alpha_k}\right|
 \frac{\max(0,h_{\Omega_-}-\delta_d)}
      {h_{\Omega_+}+\delta_d}
 &\leq
 \frac{|\beta_k\mathcal G_L(\Omega_-)|}
      {|\alpha_k\mathcal G_L(\Omega_+)|}
 \nonumber\\
 &\leq
 \left|\frac{\beta_k}{\alpha_k}\right|
 \frac{h_{\Omega_-}+\delta_d}{h_{\Omega_+}-\delta_d}.
 \label{eq:cross-ratio-bounds}
\end{align}
These inequalities follow from the triangle and reverse triangle
inequalities. They show that the large mixing in
Eq.~\eqref{eq:cross-leakage} survives at finite $d/s$ once the endpoint
correction is sufficiently small.

For the example in Eq.~\eqref{eq:cross-leakage}, already $d/s=4$
gives $\delta_d<10^{-3}$. Indeed,
$\delta_d\leq\ee^{-8}/(4\sqrt{2\pi})$,
$\ee^8\geq1+8+8^2/2+8^3/6>125$, and $\sqrt{2\pi}>2$.
At $s=s_+$, Eq.~\eqref{eq:cross-ratio-bounds} therefore gives
\[
 \frac{|\beta_k\mathcal G_L(\Omega_-)|}
      {|\alpha_k\mathcal G_L(\Omega_+)|}
 \geq\frac{99}{101}
 \frac{1-72/10201-1/1000}{1+1/1000}
 =\frac{91069191}{93757391}>\frac{97}{100}.
\]
The positive-frequency mixing is thus large on the physical half-line
as well as in its full-line extension.

More generally, choose a nonzero target term
$a_j\mathcal G_L(q_j)$ from Eq.~\eqref{eq:cross-W-exact}, including its
phase in $a_j$, and suppose $h_{q_j}>\delta_d$. The relative magnitude
of the sum of the other three terms is at most
\begin{equation}
 \epsilon_j=
 \frac{\sum_{\ell\ne j}|a_\ell|(h_{q_\ell}+\delta_d)}
 {|a_j|(h_{q_j}-\delta_d)}.
 \label{eq:cross-resolution-condition}
\end{equation}
Thus $\epsilon_j\ll1$ is a sufficient, explicitly testable condition
for approximate single-component isolation. It controls the measured
coefficient at a specified $(b,s)$, not the position or existence of
two maxima as functions of scale. Destructive interference may make
the actual contamination smaller than this conservative bound.

\end{document}